\documentclass[12pt]{article}
\def\id{\protect{{1 \kern-.28em {\rm l}}}}

\def\be{\begin{equation}}
\def\ee{\end{equation}}

\def\bea{\begin{eqnarray}}
\def\eea{\end{eqnarray}}

\def\p{{\partial}}

\makeatletter
\renewcommand\section{\@startsection {section}{1}{\z@}%
                                   {-3.5ex \@plus -1ex \@minus -.2ex}%
                                   {2.3ex \@plus.2ex}%
                                   {\normalfont\large\bfseries}}
\renewcommand\subsection{\@startsection{subsection}{2}{\z@}%
                                   {-3.25ex\@plus -1ex \@minus -.2ex}%
                                   {1.5ex \@plus .2ex}%
                                   {\normalfont\normalsize\bfseries}}

\makeatother

\def \foot {\footnote}
\def \bi{\bibitem}

\def \ha {{1 \over 2}}

\def \ci{\cite}
\def \N {{\mathcal N}}

\def\z{\zeta}

\def\p{\phi}

\def \del{\partial}

\def \aa {{\a'}}
\def\g{\gamma}

\def\z{\zeta}

\def\ov{\over}

\def\l {\lambda}

\def\foot{\footnote}

 \def \third {{ \textstyle {1\ov 3}}}
\def\det {\hbox{det}}
\def \ci {\cite}

\def \l  {\lambda}

\def \N {{\mathcal N}}

\def \D {\Delta}
\def \N {{\mathcal N}}

\def \bi{\bibitem}
\def \la {\label}

\def \l {\lambda}
\def\foot{\footnote}

\newcommand{\rf}[1]{(\ref{#1})}
\def \ov {\over}

\def\N{{\cal N}}

\def \ha{{1\ov 2}}

\def \no {\nonumber}

\def \del {\partial}

\def \bi{\bibitem}
\def \la {\label}

\def\foot{\footnote}

 \def \p {\phi}

\def \ov {\over}

\def \varpi {{\rm w}}

\def \ep {\epsilon}

\def \vp {\varphi}

\def \ed {\end{document}}

\def \te {\textstyle}

\def \ha {{{\textstyle{1 \ov2}}}}
\def \fo {{\textstyle{1 \ov4}}}

\def \D  {\Delta }

 \def \eqref  {\rf}

\def \ha {{\textstyle {1 \ov 2}}}

\renewcommand{\theequation}{1.\arabic{equation}}
\def \bi {\bibitem} 
 
\def \be {\bea}
\def \ee {\eea} 

\def \dn {{\hat \d}}

\def \na {\nabla}

\def \pe {\perp} \def \tri {{\te { 3 \ov 2}}}
\def \iffa {\iffalse} 

  \def \De {\Delta}

 \def \kk {{\rm k}}

\def \aa  {{\rm a}}

\def \vp {\varphi}

\def \p {\phi}  
\def \D {{\rm D}}

\def  \iffa {\iffalse}
\def \tri {\triv}

\def \ep {\epsilon} 

\def \text { }

\def \pe {\perp} 

\def \np {\newpage}
\def \py {{\te {5 \ov 2}}}

\def \dn {{\hat \Delta}}

\def \rr {{\, r}}

\begin{document}


\overfullrule=0pt
\parskip=2pt
\parindent=12pt
\headheight=0in \headsep=0in \topmargin=0in \oddsidemargin=0in

\vspace{ -3cm}
\thispagestyle{empty}
\vspace{-1cm}

\rightline{Imperial-TP-AT-2013-6}
 \def \G {\Gamma}

\begin{center}
\vspace{1cm}
{\Large\bf  
Weyl anomaly 
of   conformal  higher spins  on  six-sphere
}
\vspace{0.8cm}

 { 
 A.A. Tseytlin\footnote{Also at Lebedev  Institute, Moscow. e-mail:
tseytlin@imperial.ac.uk }
}

\vskip 0.10cm

{
\em 
\vskip 0.08cm
\vskip 0.08cm 
Blackett Laboratory, Imperial College,
London SW7 2AZ, U.K.
 }
\vspace{.32cm}
\end{center}

\vskip 0.10cm
\begin{abstract}
This  paper  is a sequel to  arXiv:1309.0785  were we computed the  Weyl anomaly $a$ 
(Euler density or  logarithmic  divergence   on $S^d$)  coefficient 
 for  higher-derivative conformal  higher spin field  in $d=4$  and shown  that it  matches the  
 expression  found   in arXiv:1306.5242  by a ``holographic''  method from a ratio  of 
  massless   higher spin determinants in AdS$_5$.
  Here we repeat the same   computation in  on 6-sphere  and   demonstrate   that the 
  result matches  again  the one  following from   AdS$_7$.
  We also discuss explicitly  similar  matching in the $d=2$ case. 
  
\end{abstract}

\def \D  {\hat\Delta}\def \dD {{\rm d}}

 \def \zr {{\z_{\rm R}}}
\def \tri  {{\te{3 \ov 2}}}
\def \N {{\cal N}}\def \K {{\rm K}} 
 \def \AdS {{AdS}}

 \def \K {{\rm K}} \def \kk {{\rm k}} 

  \def \De  {\Delta}

\newpage

\setcounter{equation}{0} 
\setcounter{footnote}{0}
\section{Introduction}

This paper  continues the investigation  \ci{me} 
of quantum conformal higher   spin (CHS)    models  \ci{ft,seg}  with  higher-derivative  flat-space action 
$\int d^d x  \, \p_s  P_s \del^{2s + d-4} \p_s$ \  ($P_s$ is transverse traceless  symmetric rank $s$ tensor
projector).   
Generalising this  action to curved background is a highly non-trivial problem, 
but as was argued in \ci{me},   at least in the  case of a conformally flat 
Einstein  background  (i.e.   (A)dS$_d$ or $S^d$), 
  the corresponding Weyl-covariant  $2s + d-4$ derivative 
 kinetic  operator  should factorize into product  of standard 2nd-derivative operators.\foot{Here  we shall always
   assume that $d$ is even ($d$ was called $D$ in \ci{me}).}
  
Explicitly, the partition  function of a conformal higher  spin $s$ 
field  on a  $d$-dimensional sphere  of unit radius  can be written as \ci{me} 
\be
 &&  Z_{s} ({S^d}) =\prod_{k=0}^{s-1} \Big({\det\, \big[-\nabla^2   + k - (s-1) ( s + d-2) \big]_{k\, \pe} \ov
 \det\, \big[-\nabla^2  + s - (k-1) ( k + d-2) \big]_{s\pe}}\Big)^{1/2}
 \no \\
&& \qquad\  \ \ \ \  \  \times 
 \prod^{-1}_{k'=-{1\ov 2}({ d-4 } )} \Big({1 \ov
 \det\, \big[-\nabla^2  + s - (k'-1) ( k'+  d -2) \big]_{s\pe}}\Big)^{1/2}
   \ ,  \la{0}
\ee
or, equivalently, as 
\be
 &&  Z_{s} ({S^d}) =\prod_{k=0}^{s-1} \Big(\det\, \big[-\nabla^2   + k - (s-1) ( s + d-2) \big]_{k\, \pe} \Big)^{1/2}
 \no \\
&& \qquad\  \ \ \ \  \  \times 
 \prod^{s-1} _{k'={-{1\ov 2}({ d-4 } )} } \Big(
 \det\, \big[-\nabla^2  + s - (k'-1) ( k'+  d -2) \big]_{s\pe}  \Big)^{-1/2}
   \ ,   \la{01}
\ee
where the 2nd-order differential operator $ (-\nabla^2  + M^2)_{k\pe}$   is   defined on transverse traceless  symmetric 
rank $k$  tensors.  The first line in \rf{0}   is 
 the  contribution of the  ``partially-massless''    modes (with residual gauge invariance  and thus ``ghost''  numerators)   while the second     corresponds to extra 
 ``massive''   modes    present for $d \not= 4$ (see \ci{me}   and refs. there).

 This representation  allows one  to compute the 
 CHS  partition function  on $S^d$  using standard (e.g., $\zeta$-function)  techniques, and, in particular,  
to find  the coefficient  of the   logarithmic UV divergence  or  the $\aa$-coefficient  of the Euler density term in  
 the   corresponding  Weyl anomaly. 
 
 Remarkably, the arguments  in \ci{gio,me}  suggest that $ Z_{s} ({S^d}) $  in \rf{0}   should   have also 
 a  ``holographic''  representation in terms  of the ratio of determinants of the  standard (second-derivative) 
 massless  higher  spin $s$ operators   with alternate boundary conditions in 
 euclidean AdS$_{d+1}$:
\be
  &&  {Z^{(-)}_{s0} ({\AdS_{d+1}})  \ov Z^{(+)}_{s0}  ({\AdS_{d+1}}) }   =  Z_s({S^d})  \ , \la{02}
 \\
 &&  Z_{s0} ({\AdS_{d+1}}) =
 \Big({\det\, \big[-\nabla^2  + (s-1) ( s + d-2) \big]_{s-1\, \pe} \ov
 \det\, \big[-\nabla^2  -  s + (s-2) ( s + d-2) \big]_{s\pe}}\Big)^{1/2} \ . \la{03}
 \ee
Here AdS$_{d+1}$   and its boundary  $S^d$   are  assumed to have unit radius.
The subscripts $\pm$   indicate  the  different  boundary conditions.\foot{These   correspond to dimensions 
$\Delta_+= s + d-2, \ \Delta_-= 2-s$ for the ``physical'' denominator and $\Delta_+= s + d-1, \ \Delta_-= 1-s$ 
for the  ``ghost''  numerator \ci{gio} (see also section 4).}

Let us note that  while motivated by the AdS/CFT \ci{gu,kle,gio}, 
the relation \rf{02} is  essentially  
``kinematical'' in nature (i.e.  it does not  rely on   any  non-renormalization  and should be true for any $d$)
belonging to a class of bulk-boundary determinant relations like the one discussed in \ci{bar}. 
One  should  thus  be able to prove  it  by  starting from the  one-loop path integral  in AdS$_{d+1}$   and 
``integrating out''  the values of the fields in the interior  points of AdS$_{d+1}$. 
As in the  scalar case \ci{har,dorn,diaz}  one should  pay special attention to regularization. 
Indeed,  the AdS$_{d+1}$   side of \rf{02}  is   IR   divergent while  the $S^d$   side  is   UV divergent. 
The  logarithm  of partition function  $Z_{s0}$ 
 on   AdS$_{d+1}$ is proportional to its volume   which for even $d$  has the following regularized value 
 \ci{dorn} (we shall keep track of  logarithmic  divergences only): 
\be \Omega(AdS_{d+1}) = {2 (-1)^{d\ov 2} \pi^{d\ov 2} \ov \G( { d+1 \ov 2}) } \ln L + ... \ . \la{04} \ee
   where $L\to \infty$ is  IR cutoff. 
The free energy  on $S^d$ of radius $\rr$  has  the following  structure
 \be  
&& F=-\ln Z = \ha \ln \det\, (-\nabla^2 + M^2) = -B_d \ln (L \rr)  +... \ , \la{05}\\
&&    B_d ={\te  { 1 \ov (4 \pi)^{d/2}}}\int d^d x \sqrt g\,  b_d =  {\te { 1 \ov (4 \pi)^{d/2} }}\Omega(S^d)\,    b_d  \ , \ \ \ \ \ \ 
\Omega(S^d) =  { 2 \pi^{d+1\ov 2} \ov \Gamma({ d+1\ov 2})}
 \ ,\la{06}  \ee
 where $B_d$  is the  integrated Seeley   coefficient (often called  also $a_{d/2}$) 
 of the operator $-\nabla^2 + M^2$  and $L\to \infty $  is   UV  (heat kernel) cutoff. 
 In the case when the classical theory is conformally invariant $B_d$ represents  the integrated Weyl 
 anomaly (see \ci{duf,bon,des,bol} and refs. there). 
 The total coefficient  of $\ln L$ term in $\ln Z_s$  can be found by summing  the $B_d$-coefficients 
 for the operators in \rf{01}.
 
 Identifying the IR  cutoff in the  AdS$_{d+1}$   bulk and the UV  cutoff at the $S^d$ boundary 
 the first check  of \rf{02}  is the matching of the coefficients of  the $\ln L$ terms.
 Following \ci{gio}  let us  call   $a_s$ the coefficient of the IR singular term in the  AdS$_{d+1}$   free   energy 
 in \rf{02}.   
   Comparing to the $S^d$ expression \rf{05}   we should get\foot{The minus  sign  in the relation   
 between $ B_d^{(s)}$    and  $ a_s $  is  due to the canonical 
  minus sign in \rf{05}  or the definition  of $a_s$ in \ci{gio} 
   so that it has the same sign as   the  $\aa$-coefficient in the trace anomaly.
   It is also   sensitive to the order  of the 
  signs or the power of  the  l.h.s. of \rf{02}.}   
   \be 
 B_d^{(s)} = - a_s   \  .  \la{uh} \ee
 Equivalently,   $a_s$    should be the coefficient of  the $\ln \rr$ term in free energy on $S^d$. 
 
 In the case of $d=4$ the coefficient of the 
IR divergent term     in the    l.h.s.  of \rf{02}  was  found in   to be  \ci{gio} 
\be \la{07} 
a_{s} = \te  { 1 \ov 180} \nu_s^2 (   14 \nu_s + 3 ) \ , \ \ \ \ \ \ \ \ \ \  \  \nu_s = s (s+1) \ .
\ee 
The same   expression  was   also   obtained    directly    from  
the  spin $s$ 
 CHS partition  function \rf{0}  on $S^4$  as (minus)  the   value of the total
 Weyl anomaly  coefficient $B_4^{(s)}$ \ci{me}
 \be 
 &&\te   b_4 = - \aa_s  R^*R^* =  - 24 \aa_s \ , \ \ \ \ \ \  \ \  \ \ 
 B_4^{(s)}   = { 1 \ov (4\pi)^2} { 8 \pi^2 \ov 3}    \, b_4  =   - 4 \aa_s  = - a_s  
 \ ,\la{08} \\
&&   \ \ \ \ \ \ \     B_4^{(s)} = - a_s = \te - { s^2 (s+1)^2  \ov 180}   \big(  14 s^2 + 14 s   + 3 \big)  \ .  \la{088}
  \ee 
   Our aim  here  will be  to  perform a  further  non-trivial test of  the relation \rf{02}   by considering 
   the  $d=6$    case  (and also the $d=2$ case, see Appendix). 
 The case of $d=6$  is  of    interest    in view of  the 
  AdS$_7$/CFT$_6$  duality    and also because 
 the structure  of the CHS  partition function \rf{0} changes for $d\not=4$. 
 We shall   first consider   the r.h.s. of \rf{02}, i.e.  find    the  coefficient $B_6$ \rf{06}  of logarithmically divergent term in 
 $F=-\ln Z_s$ in \rf{01}   on $S^6$. 
  
  In general, the local Weyl anomaly coefficient   has the  following structure in $d=6$
  \ci{bon,des,bas}\foot{In contrast to \ci{bas} here we do not include $ { 1 \ov (4\pi)^{d/2}}$ in the definition of $b_d$.}
 \be 
b_6 = \aa  E_6  +  \sum_{i=1}^3  c_i I_i   + \nabla_m J^m \ , \ \ \ \ \   \qquad  E_6= - \ep_6 \ep_6  RRR \ , \la{09}
\ee 
where $I_1\sim C ( \nabla^2 + ...) C, \ I_{2,3} \sim CCC  $  contain  powers  of the Weyl tensor $C$.  
Then  for a unit-radius  sphere $S^6$ \foot{Here $R_{mnkl}= g_{mk} g_{nl} - g_{ml} g_{nk}, \  \ \ R= d(d-1) = 30 $.
The (minus)   Euler density  $E_6$  is equal to $-{ 16 \ov 75} R^3$  on  a conformally flat background.}
\be  b_6 (S^6)= \aa   E_6  =   -  { 8!\ov 7}  \, \aa    \ , \ \ \ \ \ \ \    \ \ \  
 B_6(S^6)  = \te  { 1 \ov (4\pi)^3} { 16\pi^3 \ov 15}     b_6  =   { 1 \ov 60}  b_6 =   -   96 \aa   \equiv  - a     
    \ . \la{001} \ee 
For a conformally  coupled scalar  ($\dn=-\nabla^2 + { d-2 \ov 4 (d-1)} R$) 
\be \la{002} 
 \te  \aa_0= - {5 \ov  9!} \ , \ \ \ \ \ \ \ \   \ \ \ \ \ \ \ \ \ \    B^{(0)}_6  = -a_0 = { 1 \ov 756} \ . 
\ee
As we  shall find  below, for  a conformal higher spin  field in $d=6$  the total value of $B_6$   corresponding to \rf{01} 
  (generalizing  \rf{002} to any $s \geq  0$) is
\be 
&&
  B^{(s)}_6 =- a_s = \te   -\frac{(s+1)^2 (s+2)^2}{151200} 
 \left(22 s^6+198 s^5+671 s^4+1056 s^3+733 s^2+120 s-50\right)\ \no  \\
 && \ \ \ \ \ \  \te  = - \frac{1}{18900}{\nu}_s \left(88 {\nu^{3/2}_s}-110 {\nu}_s-4 {\nu}^{1/2}_s  +1\right)
\ , \ \ \ \ \  \qquad \nu_s = \fo (s+1)^2 (s+2)^2   \ .
 \la{111}
\ee
Like  in the $d=4$ expression  \rf{07}  here $\nu_s$   stands for the number  of dynamical degrees 
of a spin $s$  CHS field in $d=6$. 
Specialising  the general expression \ci{gio} for the  coefficient of the IR divergent part of the AdS$_{d+1}$  side  of \rf{02} 
   to the case of $d=6$ we will   also   show that it  indeed  matches  \rf{111}  according to  \rf{uh}. 

We shall start   
in section 2 with a general  discussion of the  values of  the 
 $\z$-function and the  logarithmic   UV divergence   coefficient  $B_d$  for a 
massive  higher  spin   operator $( - \na^2  +  M^2)_{s\perp} $   on $S^d$, and then  specialise to the cases 
$d=4$ and $d=6$. 
 In section 3 we  shall   apply  the resulting 
 expression for $B_6$ to the operators  appearing in \rf{01}  to obtain eq. \rf{111}. 
 In section 4 we shall rederive \rf{111} as the coefficient of the IR divergence of the ratio  of the AdS$_7$
 massless spin $s$ partition functions in \rf{02}.
 Section 5  will contain  concluding remarks.
 In Appendix   we  shall  consider  the $d=2$  case of \rf{02}  and 
 demonstrate  explicitly  that the AdS$_3$   expression  for $a_s$     matches 
   the coefficient $B_2$ of the UV divergence   in   the $d=2$   conformal 
 higher-spin  partition   function \rf{02}, thus providing another check of \rf{02},\rf{uh}.

\def \l {\lambda} 

\renewcommand{\theequation}{2.\arabic{equation}}
 \setcounter{equation}{0}
 \section{ $\z$-function and $B_d$ coefficient for  spin $s$  operators on $S^d$    }

 To compute $B_d$ we shall use the known solution  of the 
  spectral problem  for the   2nd-order operator $\dn_{s\perp} $ 
  defined on  symmetric traceless  transverse tensors of rank $s$   on $S^d$
  \be  \D_{s\perp} (M^2)    \equiv ( - \na^2  +  M^2)_{s\perp} \ , \ \ \ \ \ \ \ \ \ \ \  \ \ \ 
  \dn_{s\perp} (\p_s)_n = \l_n (\p_s)_n \  .  \la{1}  \ee
 The  eigen-values  and their degeneracy
  are given by \ci{rub,hi,ch}\foot{For the  proof of the  general expression 
   for $\dD_n$   (which is equal to dimension of a particular  representation of $SO(d+1)$)  see \ci{ch}.}
  \be  && 
  \l_n =  (n+s)  (n+s  + d-1) - s   +   M^2 \ , \ \ \ \  \ \ \ \ \ \ \ \     n= 0,1,2,....\la{2}  \\
  &&
  \dD_n = g_s { (n+1 ) ( n + 2 s + d-2 ) (2 n + 2s  + d-1 )\  (n+ s  +  d-3)! \ov   (d-1)! \ (n + s+1 )!  }  \ ,\la{3} \\
  &&\ 
  g_s   = { ( 2 s + d - 3) \ ( s + d - 4)! \over  ( d - 3)! \ s!} \  .\la{4} 
  \ee
 Here $g_s \equiv g_s^{(d)} $
   is the number of  components  of the  symmetric traceless  transverse  rank $s$ tensor in $d$ dimensions\foot{We will use 
 superscript $(d)$   to indicate  the number of dimensions  when  necessary.} 
 \be 
 g_s \equiv N_{s\perp}= N_s - N_{s-1}    \ , \ \ \ \ \ \ 
 N_s  = { ( 2 s + d-2) ( s + d-3)! \ov ( d-2)! \ s! }  \ , \ \ \ \ \ \ \ \    g_s^{(d)} = N_s^{ (d-1)}   \ , \la{499}\ee
 where $N_s\equiv N_s^{(d)}  $   is the number of   symmetric traceless  rank $s$  tensor  components. 
The number of dynamical components of a massless   spin $s$ field  is (cf. \rf{03}) 
\be 
\mu_s = N_{s\perp} - N_{s-1\, \perp} = { ( 2 s + d-4) ( s + d-5)! \ov ( d-4)!\ s! }  \ , \ \ \ \ \ \
\mu_s^{ (d)} = g_s^{(d-1)}= N_s^{ (d-2)}  \ .  \la{389}\ee
Note also   that  the number of dynamical degrees of freedom 
of a conformal higher spin $s$ field is (cf. \rf{01})\foot{This expression was first obtained in ordinary-derivative formulation of conformal higher spin field in general dimension $d$  in \ci{met}.}
 \be 
 &&\nu_s =   [ s + \ha (d-4) ] N_{s\perp} - \sum^{s-1}_{k=0} N_{k\perp}  =
 \frac{(d-3) (2 s + d-2)  (2 s + d-4) (s + d-4)!}{2  (d-2)!\  s!}  \ ,   \la{477} \\
  && \nu_s =    \frac{ (2 s + d-2) (s + d-4)}{2  (d-2)} \ \mu_s
 \ . \la{479} \ee
   The   $\z$-function corresponding   to the operator \rf{1} is  defined by 
  \be 
  \z_{\D_{s\perp} } (z) = \sum^\infty_{n=0}  { \dD_n\ov ( \l_n )^z}  \ . \la{7} \ee
  In  general,  it is $B_d$  and  not $\z_{\dn }  (0)$ 
that governs  the scale dependence of $\log \det\, \D$ in \rf{05}. 
 Note that the definition of $\z$ we use here 
 requires  summation  over all modes, including the  zero ones.
 Then while  for  the  operator $\dn_{s}$ defined on differentially unconstrained tensors  one has  
  $\z_{\dn_{s} } (0)  = B_d[\dn_{s}]$, 
  this is not so in general for $\dn_{s\perp}$:
   $\z_{\D_{s\perp} }(0)$  turns out to be   equal to $B_d[ \dn_{s\pe}]$ in \rf{06}  for the operator \rf{1} 
  only 
up to the  contribution of the zero  modes  of the operator 
  related to the  change of variables from an  unconstrained tensor  $\p_s$ 
to its transverse part. In the case of $d=4$ and $s \leq 2$  the reason for this  was explained in \ci{fto}:\foot{The 
 difference between $B_4$ and $\z(0)$   was   pointed out  also   in \ci{cr3}.}
to define  the  operators  acting on constrained (transverse) tensors  one decomposes the 
field  into its  transverse and gradient  parts  but that introduces $\N $ additional zero modes
 of the Jacobian  of  the change of variables. Since these  modes 
  were not present for the  original unconstrained operator  one finds 
 $B_d [ \dn_{s\pe}]= \zeta_{\dn_{s\pe}} (0) - \N$. 
 
In more  detail, starting with path integral over symmetric traceless tensor $\p_s$ 
we may change the variables  to   transverse symmetric traceless rank $s$ tensor 
$\p_{s\perp} $   and   symmetric traceless rank $s-1$ tensor   $\vp_{s-1}$
\be  && \p_s = \p_{s\perp} +   \K \vp_{s-1}\ ,  \ \ \ \ \ \ \ \ \ \ \ \ \ \ \ \nabla \cdot  \p_{s\perp}=0 \ , \la{21} \\
&&\te (\K \vp_{s-1} )_{m_1 ...m_s} = \nabla_{(m_s} \vp _{m_1 ...m_{s-1})}  - { s-1 \ov 2(s-2)   + d } \, g_{(m_s m_{s-1}} \na^n  \vp _{m_1 ...m_{s-2}) n } \ . \la{20}
\ee
Then  $\det\,\, \K$  will appear as the Jacobian. 
 The zero modes of $\K$  are  rank $s-1$ conformal Killing tensors
and 
their   number  is  dimension of $(s-1,2, 0, ..., 0)$ representation of $SO(d+1,1)$
\ci{eas} 
\be 
&&\kk_{s-1, d} = ( 2 s + d - 4) ( 2 s  + d- 3) (  2 s + d- 2) \ { (s + d- 4)!\,  ( s+ d  - 3)!\ov s! \,  (s - 1)! \, d! \, (d - 2)! }   \ . \la{19}
\ee
 Thus 
 \be \la{vv}
B_d[ \dn_{s\pe}] = \z_{ \dn_{s\pe}}(0)  - \N \ , \ \ \ \ \ \  \ \ \ \ \ \  \N = {\rm  dim \, ker}  \, \K = \kk_{s-1, d}   \ . \ee
It should be noted that this subtlety is absent if one considers  instead of $S^d$  the  non-compact 
euclidean $H^d=$AdS$_d$   background: 
then the  corresponding $\z_{ \dn_{s\pe}}(0)$-function defined according to \ci{ch} matches 
 $B_d[ \dn_{s\pe}]$.\foot{Here the zero modes are non-normalizable and effectively drop out   
 of $\z_{ \dn_{s\pe}}(0)$ on $H^d$  defined as in \ci{ch}.}
 
In what follows   we shall  be  interested in the two special cases: the familiar $d=4$ case  (to compare to the results of \ci{me} which were found 
directly from the general expression for $B_4$, i.e.    without using the spectrum on $S^4$) 
and the  new $d=6$ one. 
 One finds from \rf{2}--\rf{4},\rf{19} 
  \be 
 d=4:\ \ \ \ \ \  &&
   \l_n =  n^2 + (2s+3) n +  s(s+2)    +   M^2 \ , \ \ \ \   \ \ 
   g_4(s) = 2s +1 \ ,   \  \ \ \ \ \ \ \ \qquad 
    \la{a5}\\
   &&\qquad  \ \dD_n = \te { 1 \ov 6}  g_s (n+1 )(n+ 2s +2 ) (2 n + 2s+3) \ , \la{5}
   \\
   &&\qquad  \   \kk_{s-1, 4} = \te { 1 \ov 12}   (2s+1) s^2 ( s+1)^2 \ , \ \ \   \la{119}
   \ee
    \be 
   d=6:    &&
   \l_n =  n^2 + (2s+5) n +  s(s+4)    +   M^2 \ , \ \ \ \ \
   g_s = {\te{ 1 \ov 6}} (s+1)  (s+2) (2s +3)  , \ \ \ \la{aa5} \\
   &&
   \dD_n = \te { 1 \ov 120  } g_s 
       (n+1 ) (n+s + 2 ) (n + s +3) (n +2s +4 ) (2 n + 2s+5) \ ,  \la{6}\\
       && \ \kk_{s-1, 6} =\te  { 1 \ov 4320}  (2s+3)   s (s+1)^3 (s+2)^3  (s+3)  \ . \la{219}  \ee 
 Note that in $d=6$ 
  the number of symmetric traceless  tensor components  is (see \rf{499}) \\
  $N_s= {1 \ov 12} ( s+1) (s+2)^2  (s+3)$;    the number of transverse  components  is 
  $N_{s\perp}= g_s={\te{ 1 \ov 6}} (s+1)  (s+2) (2s +3)$;   the number of dynamical degrees of freedom 
  of a massless spin $s$ field \rf{389}  is $\mu_s=(s+1)^2$;    the 
   number of dynamical degrees of freedom of a conformal spin $s$ field  \rf{479}  is 
  $\nu_s= \frac{1}{4} (s+1)^2 (s+2)^2$.
  
  Let us now consider the  computation of the corresponding  values  of $ \z_{ \dn_{s\pe}}(0) $ in $d=4$ and $d=6$.

 \subsection{$d=4$ case }

The computation of $\z_{ \dn_{s\pe}}(z)   $ in $d=4$ was discussed in   \ci{al}. 
First, we  write \rf{7} as 
\be 
\z_{ \dn_{s\pe}} (z) =\third   ( 2s + 1)  \sum^\infty_{k=s + {3\ov 2} }  { k [ k^2 - ( s + \ha)^2] \ov k^{2z} (1 -  {h^2\ov k^2})^z}  \ , \ \ \ \ \ \ \ \ \ \ \ 
h^2 = s + {\te{ 9\ov 4} }- M^2 \la{99}\ . 
\ee
Then  using that
\be \la{exp}
 \Big(1 -  {h^2\ov k^2}\Big)^{-z} = \sum^\infty_{m=0} c_m(z)  {h^{2m}\ov k^{2m}} \ , 
 \ \ \ \qquad \qquad   c_m(z) = {  (z + m-1)! \ov m! \ (z-1)!}  \ , \ee
we get 
\be 
&&\z_{ \dn_{s\pe}} (z)  =\third   ( 2s + 1)  \sum^\infty_{m=0 } c_m(z)  h^{2m}  \Big[ \zr(2z+ 2m-3 , s+ \tri) \no \\
&&\qquad \qquad \qquad\qquad\qquad \qquad \qquad \qquad \ -\ ( s + \ha)^2 \zr (2z+2m -1, s + \tri) \Big] \ ,  \ \ \
\la{91} 
\ee
where $ \zr(z , b) \equiv  \sum_{n=0}^\infty ( n + b)^{-z} $.  To find  the limit $z\to 0$ 
we need to use that    the terms  with $m=1,2$ may have  a pole as 
$\zr(x,b) = { 1 \ov x-1} - \psi (b) + ...$.  Then we end up with 
\be \te   &&\z_{ \dn_{s\pe}} (0)  =\te {  {1 \ov 3} ( 2s + 1)}  \Big[ \zr(-3 , s+ \tri) - ( s + \ha)^2 \zr (-1, s + \tri) \no\\
  && \qquad \qquad\qquad \qquad\ \ \ \ \ \ \ \ -   \fo   ( s + {\te{  9\ov 4}} - M^2) ( 2 s^2   + s   - {\te {7 \ov 4}}   + M^2) \Big] \ , \la{8}\ \ \ 
\ee
where  
$ \zr(-1 , b) =- \ha b^2 + \ha b - { 1 \ov 12}$ and $ \zr(-3 , b) =-\fo b^4 + \ha b^3 - \fo b^2  +  { 1 \ov 120}$.
Finally, 
\be   \z_{ \dn_{s\pe}}  (0) = \te
  \te { 1 \ov 180}   (2s+1)    \Big[ 15 M^4  + 30 (s^2-2) M^2    + 58   - 10 s - 70 s^2   + 15 s^4       \Big]\ .  \la{9}
\ee
Then using \rf{vv},\rf{119}   we get 
\be 
&&B_4[ \dn_{s\pe} (M^2) ]  =  \z_{ \dn_{s\pe}}  (0)   -   \kk_{s-1,4} \no \\ 
&&\qquad \qquad \qquad =   \te { 1 \ov 180}   (2s+1)    \Big[ 15 M^4  +  30 (s^2-2) M^2    + 58   - 10 s -  85  s^2    - 30  s^3       \Big] \ .    \la{93} 
\ee 
Taking into account \rf{08} this matches the expression for $\aa[ \dn_{s\pe}(M^2) ] $  which was found  \ci{me} 
directly from the standard algorithm for $B_4$ \ci{gil}  and  using  that \ci{me} 
\be 
&&   \det\,  \dn_{s\pe} (M^2) =  { \det\,  \dn_{s} (M^2) \ov   \det\,   \dn_{s-1} (M^2-  2s - d + 3  )} \ , \la{2222}\\
&&B_d [ \dn_{s\pe} (M^2)] = B_d [ \dn_{s} (M^2)]  - B_d [ \dn_{s-1} (M^2-  2s - d + 3  )]  \ . \la{22}
\ee
Applying \rf{93}   to find the total $B_4$   or $\aa$ coefficient \rf{08}  corresponding to  the  $d=4$ CHS partition \rf{0}   one  ends up 
with   \rf{088} \ci{me}.

Let us  note that  the same expression \rf{93}    can be found also by   considering instead of $S^4$ the non-compact 
 $H^4$ (euclidean AdS$_4$)   background.
  Indeed, the local  expressions for  the coefficient $b_4$ in \rf{06}  should match since 
  it  depends on  the square of the curvature while $R(S^4) = - R(H^4)$
  (one should also change the sign of the $M^2$  term as it enters as $M^2 \ep,\   R= d(d-1)\ep, \ \ \ep=\pm 1$). 
 Computing the corresponding  value of $\zeta_{ \dn_{s\pe}} (0)$ 
 as in  \ci{ch2}  (where its ``un-integrated''  value was found)  and   taking into account \ci{ch}  that the regularized volume 
 of $H^4$  is\foot{In general, for even-dimensional  case 
   one has $\Omega(H^{2n}) =\pi^{ n - {1\ov 2}} \Gamma( - n + {1\ov 2})$ \ci{dorn}.}
 $\Omega(H^4) = { 4 \pi^2 \ov 3}$ while  $  \Omega(S^4) = {8 \pi^2 \ov 3}$ 
 we conclude that  $B_4$ and $\zeta^{(H^4)}_{ \dn_{s\pe}} (0)$  should be equal up to the  factor of 2 coming
  from the ratio of the  two volumes. 
 Explicitly,  given the operator  $\dn_{s\pe} (M^2) = (-\nabla^2   +   M^2)_{s\pe}$  one finds
 \be 
  &&   B_4 [ \dn_{s} (M^2)]  =  \zeta^{(S^4)}_{ \dn_{s\pe}(M^2)} (0)   -   \kk_{s-1,4} =\ 
        2\, \zeta^{(H^4)}_{ \dn_{s\pe}(-M^2)} (0) \ , \la{23} \\
        &&
  \zeta^{(H^4)}_{ \dn_{s\pe}(-M^2)} (0) 
   =\te  { 1 \ov 24} ( 2s  + 1) \Big[h^4 -   ( s + \ha)^2  ( 2 h^2 + { 1 \ov 6}) -  { 7 \ov 240} \Big] \ , \ \ \ \ \ \ 
 h^2 = s + { 9\ov 4}  -  M^2  \ . \ \ \ \ \ \la{933} 
 \ee       
The expression \rf{933}  was 
also  used in \ci{gk}  in the  computation of the UV divergent term of the massless  higher spin theory in AdS$_4$.

In general, the partition function     \rf{03}  \ci{gab,lal}   of a   massless  spin $s$ field   in AdS$_d$  ($\ep=-1$) 
or   dS$_d$   or $S^d$ space  ($\ep=+1$)  can be written as  
\be 
Z_{s0}= \Big({ \det\, \dn_{s-1\, \pe}[M^2_{s-1,s}\ep ] \ov \det\, \dn_{s\pe}[M^2_{s,s-1}\ep ]} \Big)^{1/2} \ , \ \ \ \ \ \ \ \ 
\ \ \ \ \ \ \     M^2_{n,k} \equiv   n - (k-1) (k + d-2) \ . \la{75}
\ee
Then we may  use \rf{93} 
to find  the  coefficient \rf{05}  of  the   divergent  term in $F=-\ln Z_{s0}$  in $d=4$: 
\be 
&& d=4:\ \ \ \ \ \    M^2_{s,s-1} =- s^2 + 2s + 2   \ , \ \ \ \ \ \ \  \ \ M^2_{s-1,s} =   -s^2    + 1 \  , \la{zzz} \\
  \te &&   B_4^{(s0)}\equiv  B_4[\dn_{s\pe} (M^2_{s,s-1})] -\te  B_4[\dn_{s-1\, \pe} (M^2_{s-1,s})]  =-\te
  \frac{1}{90} \left(75 s^4-15 s^2 +2\right)
  \ . \la{yyy}
 \ee
 This is equivalent to the expression  obtained  in \ci{gk} using \rf{933}.  
  It was  found there  that the $\z$-function  regularized sum of  the   values of $ \zeta^{(H^4)}(0)$ over all  massless 
  spins $s >0$  plus 
  the  $s=0$  (scalar)  contribution 
vanishes.\foot{Note that   the $s=0$  value  of  $B_4^{(s0)}$  in \rf{yyy} 
is {\it not }  equal  to the  conformal scalar contribution $- { 1 \ov 90}$ but is twice this value (the reason is that here 
the  ``ghost''  contribution $-B_4[\dn_{s-1\, \pe} (M^2_{s-1,s})]$ does not vanish for $s=0$ and effectively doubles the
``physical''  mode contribution). 
Regularizing the  sum  $\sum_{s=1}^\infty   B_4^{(s0)}$  with $\z$-function gives $ - { 2 \ov 90} \z(0)= { 1\ov 90}$ 
which cancels against the  separate massless scalar contribution \ci{gio}.}
Let us note that the  same conclusion  applies  also  for   the corresponding  values of  the massless   higher spin 
 $\z$-function 
  computed  on  $S^4$ or  dS$_4$:
the sum over the  zero-mode terms in \rf{23},\rf{119}   given by (cf. \rf{03}) 
$ \sum_{s=1}^\infty ( \kk_{s-1, 4} - \kk_{s-2, 4}) =  \sum_{s=1}^\infty  \, \te  { 1 \ov 6}  (s^2 +  5 s^4) $
vanishes separately when $\z$-function  regularized. 

 \subsection{$d=6$ case }

According to  \rf{vv}  we should have  the following relation between $B_6$ in \rf{05} and the corresponding 
$\zeta$-function on $S^6$
\be \la{31}
B_6 [\dn_{s\pe}] = \z_{\dn_{s\pe} }(0)  - \kk_{s-1, 6}   \ , \ee
where $ \kk_{s-1, 6}$  is given in \rf{219}. 
The computation of   the $\z_{\dn_{s\pe} }(0) $ in $d=6$  uses \rf{aa5},\rf{6}   and follows the same lines as in $d=4$. 
The counterpart of \rf{99} is 
\be 
\z_{\dn_{s\pe} } (z) = {\te {1\ov 60}}  g_s \sum_{k=s + {5\ov 2} }^\infty  {k (  k^2-{1\ov 4}  ) [ k^2 - ( s+{3\ov 2})^2]  \ov  (k^2-h^2 )^{z}} \ ,\qquad  \ \ \ \ \ \
\te h^2= s + { 25\ov 4} - M^2 \ ,  \la{32}
\ee
 and  using \rf{exp}
we get 
\be
&&  \z_{\dn_{s\pe}}(z) = {\te {1\ov 60}}  g_s
    \sum^\infty_{m=0 } c_m(z)  h^{2m}  \Big[   \zr(2z+ 2m-5 , s+ \py)   \no\\ && \qquad \qquad 
    -  ( s^2 + 3 s  + \py  )  \zr(2z+ 2m-3 , s+ \py)  
       + \fo   ( s + \tri )^2 \zr (2z+2m -1, s + \py)  
        \Big] \ \ \  \ \ \la{33} 
\ee
Taking the limit $z\to 0$  gives (cf. \rf{8})
\be 
&&\z_{\dn_{s\pe} }(0) = {\te {1\ov 60}}  g_s
     \Big[   \zr(-5 , s+ \py)   -    ( s^2 + 3 s  + \py  )      \zr(-3 , s+ \py) 
            + \fo   ( s + \tri   )^2 \zr ( -1, s + \py)    \no\\
    && \ \ \ \ \ \qquad \qquad \qquad \qquad   \te   +   { 1 \ov 6}  h^6   -     { 1 \ov  4}   ( s^2 + 3 s  + \py    ) h^4   
    +    { 1 \ov 8}    ( s + \tri   )^2  h^2 
        \Big] 
\la{34} 
\ee
As a result, 
\be\te  && \te  \z_{\dn_{s\pe} }(0) = 
\frac{(s+1) (s+2) (2 s+3)}{453600}
 \Big[ -210 M^6-315 M^4 (s^2+s-10)+630 M^2 (s^3 + 8 s^2 + 8 s -24)\no 
 \\
 &&  \qquad \qquad     \qquad \qquad 
 + 22780 - 17514 s - 15288 s^2 + 840 s^3 + 2940 s^4 + 945 s^5 + 105 s^6 \Big]\la{35}
\ee
Then eq.\rf{31}   implies   that (cf. \rf{93}) 
\be && \te     B_6[\dn_{s\pe} (M^2)] = 
\frac{(s+1) (s+2) (2 s+3)}{453600}
 \Big[ -210 M^6-315 M^4 \left(s^2+s-10\right)\no \\  && 
 + 630 M^2 (s+6) (s^2 + 2s -4)  
  +  22780  - 18774 s - 19488 s^2 - 4515 s^3 - 315 s^4   \Big]\ .    \la{36}
\ee
In particular, in  the case of the conformal scalar $s=0, \ M^2 = {d-2\ov 4 (d-1)} R = \fo d (d-2) =6$ we get 
$B_6= { 1 \ov 756}$, i.e. the standard value \rf{002}. 

It  should be   possible of course to   find   \rf{36}  directly from the 
general expression \ci{gil}  for the $b_6$  heat kernel coefficient of a  2nd-order   differential operator  in  curved space, 
but  in the  arbitrary spin $s$ case in $d=6$ 
this  computation appears to be more  involved than  the one 
 based on  the  $\z$-function on $S^6$ 
 presented here.

\renewcommand{\theequation}{3.\arabic{equation}}
 \setcounter{equation}{0}
 \section{$B^{(s)}_6$ coefficient   in  conformal   spin $s$  partition  function on $S^6$ }

Let us now apply the general  expression \rf{36} to  find the $B_6^{(s)}$   coefficient corresponding to the CHS partition function \rf{01} 
on $S^6$. 
Explicitly, in $d=6$ we get 
\be
  Z_{s}({S^6}) =\prod_{k=0}^{s-1} \Big[ \det \dn_{k\perp} ( M^2_{k,s} ) \Big]^{1/2}
 \prod^{s-1}_{k'= -1}  \Big[ \det \dn_{k\perp} ( M^2_{s,k'} ) \Big]^{-1/2}   
   \ , \ \ \     M^2_{k,m}  =   k - (m-1) ( m + 4).\ \    \la{zz}
\ee
Using \rf{36}  we find   for the 
  total anomaly coefficient (cf. \rf{088})
\be 
&&   B^{(s)}_6 = \sum_{k'=-1}^{s-1}  B_6[\dn_{s\pe} \big(  s - (k'-1) ( k'+  4)  \big )]
-  \sum_{k=0}^{s-1}  B_6[\dn_{k\pe} \big(  k - (s-1) ( s+  4)   \big)]\no \\
&& \qquad=
\te -\frac{(s+1)^2 (s+2)^2}{151200} 
 \left(22 s^6+198 s^5+671 s^4+1056 s^3+733 s^2+120 s-50\right)\ . \la{40}
\ee
 Let us note that the $k=s-1,\ k'=s-1$ terms in \rf{zz}    represent the  partition function of massless
    spin $s$ field  on $S^6$  (or dS$_6$)  which is the same   as the AdS$_6$ one in  \rf{03}
   up to the sign of the  dimensionless mass  parameters:  on $S^6$   we have 
   \be  
M^2_{s,s-1} =  - s^2 + 6 \ , \ \ \ \ \ \ \ \ \
M^2_{s-1,s} =  -s^2 - 2s + 3\ .    \la{37}
 \ee
 We  find  for the contribution of  this massless    spin $s$ factor  (cf. \rf{yyy})
 \be 
  \te &&   B^{(s0)}_6 = B_6[\dn_{s\pe} (M^2_{s,s-1})] -\te  B_6[\dn_{s-1\, \pe} (M^2_{s-1,s})] \no \\
  &&\qquad \  =\te
 - \frac{(s+1)^2 }{15120} 
  \left(63 s^6  +  378 s^5   + 847 s^4  + 868 s^3  + 378 s^2 + 28s  -    20  \right)\ .  \la{38}
 \ee
 For $s=0$  this   equals to the  conformal scalar   value \rf{002} 
  as in this case $B_6[\dn_{s-1\, \pe} (M^2_{s-1,s})]$   vanishes.

\renewcommand{\theequation}{4.\arabic{equation}}
 \setcounter{equation}{0}
 \section{$a_s$ coefficient  in  ratio of    massless  spin $s$ partition functions in   AdS$_{7}$}

Let us now   show that exactly the same expression \rf{40}   appears  as a coefficient of the 
 IR divergent term      in the ratio of the massless   spin $s$ partition   functions in AdS$_7$  
 in the l.h.s. of  eq.\rf{02}. 
We shall  first  review the  general   expression   for  this    coefficient 
 found in   \ci{gio}   and  then apply  it  to the case of $d=6$.
 
Starting with a mass  $m$   spin $s$  operator  in AdS$_{d+1}$  of  unit radius ($  \ep=-1$) 
 \be \la{43} 
 \dn(M^2)_{s\perp}  = (- \nabla^2    + M^2 \ep)_{s\perp} \ , \ \ \ \ \ \ \ \ \ \ \ \ \ \
 M^2 =  - m^2   +  s - (s-2) (s + d-2)  \ , \ee 
one     finds that  the powers  of  near-boundary  asymptotics  of  the  corresponding   solutions are $\g_\pm = \Delta_\pm -s$ 
where \ci{mmm}
 \be 
&& \Delta_\pm (m)  = {\ha d } \pm \sqrt{   m^2   + ( s+ \ha d - 2)^2 } 
= {\ha d } \pm \sqrt{  \fo d^2 + 3 s - 4 - M^2}   \ ,  \la{44}\ \\
 &&
 \De_+ \equiv  \De_+(0) = s + d -2 \ , \ \ \ \ \ \  \De_- \equiv 
  \De_-(0) = 2-s   \ , \ \ \ \ \ \ \ \     \De_- = d-\Delta_+  \ .  \la{45}\
\ee
These  $\De_{\pm}$  apply  to   the physical (spin $s$) part of \rf{03} 
while for  the   ``ghost''  (spin $s-1$) part of \rf{03}
$\Delta'_+ = s + d-1, \  \Delta'_- = 1-s$ \ci{gio}.
As discussed  in the Introduction, 
 the partition function of a constant-mass operator  on  AdS$_{d+1}$ is
  proportional to its volume   which for even $d$ is   IR divergent (see \rf{04}). 
Calling  the  coefficient  of the $\ln L$ term in the corresponding free energy
 $F =\ha \ln \det\,   \dn_{s\perp}(M^2)  $   as $a_s(\Delta)$ where $\Delta= \Delta_+$  in \rf{44}   one finds that   \ci{gio}  
 \be 
&& \delta  a_{s} (\Delta)  \equiv  a_s(\Delta)  -  a_s( d -\Delta) \no \\ 
&&= 
 - { 2 g_s^{(d+1)} \ov \pi d!} \int^\Delta_{{1\ov 2}  d} dx\ (x- \ha d) ( x+ s-1) (x- s-d +1) \G( x-1) \G( d-1-x) \sin ( \pi x )\ , \ \ \la{46}\ \ \ \ \ \ 
\ee
where $g_s^{(d+1)}$ is  the same as $g_s$  in   \rf{4}  with $d \to d+1$. 
 Then the coefficient 
 $a_s$ corresponding  to the  ratio of the partition functions appearing in the l.h.s. of \rf{02} 
 can be found as  
\be 
a_{s}= \delta  a_{s} (\Delta_+ ) -  \delta  a_{s-1} (\Delta_+' )
=   \delta  a_{s} ( s + d-2  ) -  \delta  a_{s-1} (  s + d-1   )     \ . \la{47}   
\ee
The special cases of $d=2$  and $d=4$ were  already discussed in \ci{gio}.  Doing   the integral in \rf{46} gives\foot{In  
 $d=2$  one finds  from \rf{4}  that  $g_s^{(d+1)}=2$ for $s \geq 1$  and 1 for $s=0$.}
\be 
&& d=2: \ \ \ \ \ \   
\delta  a_{s} (\De)  = \te  { 2 \ov 3}   (\De-1) \Big[ 3 s^2     -   (\De-1)^2\Big] \ , \ \ \  \la{493}
\\
&& d=4: \ \ \ \ \ \  \delta  a_{s} (\De)  = 
\te \frac{(s+1)^2}{180}  (\De-2)^3 \Big[5 (s+1)^2     -  3 (\De-2)^2\Big] \ . \ \ \  \la{49}
\ee
Using these  expressions in \rf{47}    leads to  (here for $d=2$  $s \geq 2$      and $a_0= {1\ov 3}, \ a_1= {1\ov 3}$)
\be 
&& d=2: \ \ \ \ \ \   
  a_{s}    = \te  { 2 \ov 3}     + 4 s (s-1)      \ , \ \ \  \la{494}
\\
&& d=4: \ \ \ \ \ \   a_{s}    =  \te  { s^2 (s+1)^2  \ov 180}   \big(  14 s^2 + 14 s   + 3 \big)
 \ . \ \ \  \la{49x}
\ee
Thus in $d=4$  one finds  $a_s $  in \rf{07} that matches $B_4^{(s)}$ \rf{088} derived  in \ci{me} directly 
 from \rf{0} (see also section 2.1).

 The $d=2$    coefficient  \rf{494}   (rescaled by   $ - {  3}$)  was  interpreted    in \ci{gio}
 as  the central charge  $c_s= -2 [ 1 + 6 s (s-1)]$  ($s\geq 2$) of the  
  first-order   bc-ghost system   with  weights   $s$ and $1-s$
 corresponding to  spin $s$  W-gravity field   \ci{mats,hul,pop}. 
  In Appendix   we  shall demonstrate    that the AdS$_3$   prediction  \rf{494}     matches
   the  $B_2$   anomaly  coefficient  for   the  $d=2$ case of  the 
    conformal   higher spin  partition   function \rf{01}.

Let us  now  consider
   the  $d=6$ case.  Computing the integral in \rf{46} we get  (cf. \rf{493},\rf{49}) 
 \be 
 \delta  a_{s} (\Delta)  = 
\te \frac{(s+1) (s+2)^2 (s+3)}{453600}  (\Delta-3)^3 \Big[- 35 (s+2)^2   +  21[ (s+2)^2+1] (\Delta-3)^2  -   15 (\Delta-3)^4\Big]    \la{48}
\ee
Let us recall  again   that the  normalization of $a_{s}$ in \rf{46} 
 is such that it is the coefficient of the logarithm of the radius  of $S^d$, i.e. 
 it is 
 equal to  minus  the corresponding value of $B_d$:  
 in the case of $d=6$ 
  for  $s=0,\  \De={d\ov 2} + 1 = 4$  eq.\rf{48} gives $- {1 \ov 756}=-B^{(0)}_6$  (cf.  \rf{002}). 
  
   Applying \rf{48} to the case of \rf{47}   we  find 
\be 
 d=6: \ \ \ \ \ \ a_{s}&=&   \delta  a_{s} ( s + 4  ) -  \delta  a_{s-1} (  s + 5   )\qquad \no \\
 &=&
  \te  \frac{(s+1)^2 (s+2)^2}{151200}
   \left(22 s^6+198 s^5+671 s^4+1056 s^3+733 s^2+120 s-50\right)  \la{50}
\ee
This  is the same expression as in \rf{111}, i.e.  it  matches  the expression 
 \rf{40}  for $-B^{(s)}_6$ found     above directly from the CHS partition function on $S^6$.

\renewcommand{\theequation}{5.\arabic{equation}}
 \setcounter{equation}{0}
 \section{Concluding remarks}

To summarize,  in this paper we  have  shown   the 
  agreement \rf{uh}  between the UV divergence  coefficient $B^{(s)}_6$  \rf{40}  of the conformal higher spin partition function 
  on  $S^6$   and the IR 
divergence  coefficient $a_s$   in the ratio of massless  higher spin partition functions with alternate boundary conditions 
on  AdS$_7$. Together  with the corresponding $d=4$ results of \ci{gio,me}  this  
provides a    non-trivial  test  of the relation \rf{02}.  We  also demonstrate  a similar  matching in the $d=2$ case in Appendix below.

In $d=4$  the sum of  the   anomaly  coefficients  $a_s$   in   \rf{07} over all  spins $s$ 
vanishes \ci{gio}  when  computed using  the standard $\z$-function  prescription. 
The same  is true  also for the sum of the $s \geq 1$  massless   spin $s$  divergence coefficients 
in  \rf{yyy}   plus  the $s=0$ conformal scalar contribution \ci{gk}. 
 In  the $d=6$ case we discussed here  the corresponding sums  of the coefficients in 
 \rf{40}   and  in  \rf{38}  do not appear to vanish.   This may not be surprising  since 
 in $d=6$   there  is   no   a priori reason to sum 
 over  all spins with weight one and, moreover, to consider only totally symmetric  traceless tensor representation.\foot{For example, 
  one may  include  also the  self-dual 2-form field which contributes  ${221\ov 210}$
 to  the Weyl anomaly coefficient  $B_6$ on $S^6$ 
  \ci{bas}.}

In general, it would be interesting also   to study the 
$d=6$  conformal higher spin partition function   on  other backgrounds, e.g., on Ricci-flat one  as in $d=4$ case  in \ci{me}. 
The corresponding   covariant  and Weyl-invariant  CHS action should have the structure 
$\int d^6 x \sqrt g\,  \p_s ( \nabla^{2s +2}  + ...) \p_s = \int d^6 x \sqrt g\,     C_{2s} ( \nabla^2 + ...)  C_{2s}$
where  the rank $2s$ tensor $C_{2s}$  is a gauge-covariant CHS field  strength  $C_{2s} \sim  P_s  \nabla^s \p_s+...$. 
This action is known explicitly only  for lowest values of the spin. For $s=1$  the field strength $C_2$ is   the antisymmetric tensor 
and the  2nd order Weyl-covariant  operator $( \nabla^2 + ...) $  acting on it can be found, e.g.,  in \ci{er}. 
For $s=2$  the field strength $C_4$  is  the same as   the  Weyl tensor  and the   corresponding   Weyl-covariant 
operator $( \nabla^2 + ...) $ is the same that appears in  the $I_1\sim C ( \nabla^2+...) C $ term in the trace anomaly \rf{09}  \ci{bon}  (see also  \ci{er}). 
The ``minimal''  $d=6$ Weyl gravity action $\int d^6 x \sqrt g\,  I_1 $ 
(which can be expressed in terms of Ricci tensor  $I_1 \sim  R_{ab}  ( \nabla^2+...) R_{mn}$ \ci{bon})
 admits an equivalent  representation \ci{mets}
in terms of a collection of fields  with ordinary  (2nd-derivative)  kinetic terms.\foot{As was shown in \ci{mets}, 
the other two Weyl invariants $I_2,I_3 \sim CCC$  may also   be expressed in terms of  fields of  the ``ordinary-derivative'' formulation
 but  that   leads to  higher than second derivative terms and that may be   considered as an argument for   $I_1$ 
 as the  natural   conformal  spin 2 action in $d=6$    provided   one  uses  the  
 ordinary-derivative formulation of \ci{mets} as a starting point.} 
Such   an ordinary-derivative    description of the CHS field  with any spin $s$ and in any  even  dimension $d$  
is known in flat space \ci{met},     and,   following the  $s=2$  example  \ci{mets},    it 
may serve as a starting point  for constructing  a covariant CHS $s \geq 2$ actions
in generic curved backgrounds.

\iffa 
Let us now  see  what happens   if we formally sum over  all  spins with equal weight
 in  the $d=6$  case. 
From \rf{40} we find 
\be \la{41} 
  \sum_{s=1}^\infty  B^{(s)}_6
       =\te \frac{\zeta (0)}{756}+\frac{\zeta (-1)}{1260}-\frac{17 \zeta (-3)}{180}-\frac{43 \zeta (-5)}{240} 
-\frac{11 \zeta (-7)}{210} -\frac{11 \zeta (-9)}{5040} = -\frac{19}{18900}
 \ . \ee
 Adding   the $d=6$   conformal  scalar $s=0$ contribution, or, equivalently,  considering the sum in \rf{41} 
 starting from $s=0$ we also get  a  non-zero result 
 \be \la{42} 
  \sum_{s=0}^\infty  B^{(s)}_6 = \te \frac{1}{756}   -\frac{19}{18900} = \frac{1}{3150} \ . \ee
  Since this number is smaller than  any other standard field contributions\foot{For example, 
  the Weyl anomaly coefficient  $B_4$ corresponding to the self-dual 2-form field on $S^6$  is ${221\ov 210}$
  \ci{bas}.}
  it  is not clear  how it   can be cancelled out. 
  
  If we consider  just the massless   spin $s$  contribution to  \rf{40} 
  (which  corresponds   to the $k=k'=s-1$ terms in the  sums in \rf{40})  given in \rf{38}   we get 
  \be 
  \sum_{s=1}^\infty  B^{(s0)}_6 
 ={ \te  \frac{\zeta (0)}{756}     +\frac{ \zeta (-1)}{1260}    -\frac{59 \zeta (-3)}{540} -\frac{7 \zeta (-5)}{36}
   -\frac{\zeta (-7)}{30}  = -{19\ov 18900}} \ , \ \ \ \ \   \sum_{s=0}^\infty  B^{(s0)}_6  =\te   \frac{1}{3150} \ .
  \la{39}
 \ee 
 This  sum is thus   also non-zero  for massless spins in AdS$_6$, in   contrast  to what   was found in 
 AdS$_4$ case \ci{gk}. 
 
 Comparing \rf{41},\rf{42}  and \rf{39}   we conclude that the non-vanishing of the sum of the conformal higher 
 spin     anomaly  coefficients   is solely due to the  
  massless   spin $s$  contribution to the CHS partition function \rf{0}.
 It  remains to understand    if this is  just a coincidence or has  some useful  interpretation. 
 \fi 

 
\section*{Acknowledgments}
We are  grateful  to   A. Barvinsky,  R. Metsaev,   R. Roiban,  E. Skvortsov   and M. Vasiliev  for  useful  discussions.
We thank I. Klebanov for useful  comments on the draft  and pointing out 
that the AdS$_7$ expression  \rf{50} was independently found also  in \ci{saf}. 
This  work was supported by the ERC Advanced grant No.290456
and also by the STFC grant ST/J000353/1.

\np
 \appendix

\section*{Appendix:  \\   Partition function and    $B_2$ coefficient  of   conformal higher spins  
on $S^2$}

\refstepcounter{section}
\def\theequation{A.\arabic{equation}}
\setcounter{equation}{0}

Here we shall show  that the AdS$_3$   prediction for $a_s$    \rf{493}  is  indeed    the same  \rf{uh} as   
the   logarithmic  UV  divergence coefficient   $B_2$    in   the   conformal 
 higher-spin  partition   function \rf{0}  specialised to   the   $d=2$ case. 
 
 Naively, the $d=2$   limit    of the  conformal higher spin 
   action   should start with  a  $\del^{2s+d-4} = \del^{2s-2}$ term 
 ($s\geq 2$).   However,   the  $d=2$   case  of the  CHS  theory is  special -- 
 here the number of  components  $N_s$  \rf{499} of a symmetric traceless   rank $s$ tensor 
   is $s$-independent:   \ $N_s= 2$   for $s \geq 1$  ($N_s=1$ for $s=0$). 
   Then the number of the corresponding transverse components $g_s=N_{s\perp}$ \rf{499}  vanishes  for $s\geq 2$:
   $N_{s\perp}= N_s - N_{s-1} =0$    ($N_{1\perp}=1$).  Equivalently,    a symmetric rank  tensor CHS field 
   $\p_s$    can be  completely gauged away by  a combination of the  gradient  gauge symmetry 
   (generalized reparametrizations)  and the algebraic  gauge symmetry (generalized  Weyl symmetry), 
   i.e. there is no non-trivial  gauge-invariant  field strength $C_{2s}\sim  P_s \del^s \p_s $  (this is an $s \geq 3$ 
    generalization of the fact of the 
  absence of  Weyl tensor in $d=2$).  
    
    Thus  the  classical $d=2$ CHS action is trivial  (a  familiar fact for $s=2$ or  gravity in $d=2$). Still, 
    non-zero   contributions to the corresponding partition function   may come  from the gauge-fixing  or  ghost sector.
    Indeed, the   number of dynamical degrees of freedom  of a  CHS   field in $d=2$ as  following from 
    the  general expression in 
    \rf{477} is $\nu_s=-2$
 (again, a  well-known  result for  $d=2$ gravity with trivial  Einstein term  action). 
 More precisely, the CHS action in the  path integral  for the  partition function in a background 
 covariant  harmonic gauge $(\nabla\cdot  \p_s=0$) 
 will    have  actually a non-trivial   $\p_s \del^{2s-2} \p_s + ...$   kinetic term but  it will  be 
 coming solely from the  gauge-fixing term.\foot{For 
 example, 
 for $s=2$  the two   components of the  traceless rank  2 tensor $\p_2$ (or  $h_{mn}$  fluctuation of  metric)   will   enter as 
 $ (\nabla^m h_{mn})^2 \sim  h_{++} \nabla^2 h_{--} + ...$.}
 Thus, despite the triviality of the  classical gauge-invariant CHS action,  the corresponding 
 partition function will still 
 contain    ``physical''   determinants of  spin $s$ operators   coming  from the  gauge-fixing term. 
 
 Indeed, the   $d=2$    limit of  the CHS partition function \rf{01}  is  found to be 
  \be
 Z_{s} ({S^2}) =\prod_{k=0}^{s-1} \Big[\det \dn_{k \pe} ( k - s(s-1))\Big]^{1/2}
 \prod^{s-1} _{k'=1 } \Big[ \det   \dn_{s \pe} ( s - k'(k'-1)) \Big]^{-1/2}  \ . 
     \la{012}
\ee
Using \rf{2222} this may be written explicitly  in terms of unconstrained   operators as ($\dn_{-1}\equiv 1$) 
\be 
 Z_{s} ({S^2}) =\prod_{k=0}^{s-1} \Big[       {   \det\,  \dn_{k} ( k - s^2 + s  )   \ov   \det\,  \dn_{k-1} ( k - s^2 - s +1 ) }         \Big]^{1/2}
 \prod^{s-1} _{k'=1 } \Big[{ \det\,  \dn_{s-1 } (1- s - k'^2 + k' )   \ov  \det\,    \dn_{s } ( s - k'^2 + k' )    }   \Big]^{ 1/2}  \ . 
     \la{0122}
\ee
 Given an operator $\dn_k(M^2)= - \nabla^2 + M^2$  defined  on unconstrained  symmetric traceless   rank $k$ tensor 
  the corresponding Seeley coefficient \rf{06}  in the free energy \rf{05} 
 on  unit-radius $S^2$  (with curvature $R= d(d-1) = 2$) is 
 \be 
 && B_2[ \dn_{k} (M^2)]  = \te { 1 \ov 4 \pi} \Omega(S^2)\,   b_2 = \,  b_2  \ , \ \ \ \ \ \ \ \ \ 
 b_2  = N_k (   { 1\ov 6}  R - M^2 )\ , \ \ \ \la{779} \\
 &&
 k\geq 1: \ \ \ \  b_2= 2 \te  ( {1 \ov 3}  - M^2 )  \ , \ \ \ \ \ \ \ \ \ \ \ 
k=0: \ \ \ \   b_2=   {1 \ov 3}  - M^2 
 \ .  \la{777}\ee
 Applying \rf{22}   we find 
 \be 
&& k \geq 2: \ \ \ \   
B_2 [ \dn_{k\pe} (M^2)] = B_2 [ \dn_{k} (M^2)]  - B_2 [ \dn_{k-1} (M^2- 2k +1 )]  
 = - 4k + 2 \ ,  \la{770}  \\
&&    
B_2 [ \dn_{1\pe} (M^2)] = \te - { 2 \ov 3} - M^2  \ ,  \ \ \ \ \ \ \ \ 
  \ \ \ \   B_2 [ \dn_{0\pe } (M^2)]  =B_2 [ \dn_{0 } (M^2)]=    \te  {1 \ov 3}  - M^2   \ .\la{771}
  \ee
 Then the total  $B_2$ coefficient  in free energy \rf{05}  corresponding to     \rf{012}   is 
  ($s\geq 2$) 
\be 
B_2^{(s)}&=& \sum_{k'=1}^{s-1} B_2 [ \dn_{s\pe} ( s - k'(k'-1) )] - \sum_{k=0}^{s-1} B_2 [ \dn_{k\pe } ( k - s(s-1) )]  \no \\
&=&   B_2 [ \dn_{s\pe} ( s  )]  - B_2 [ \dn_{1\pe } ( 1- s(s-1) )]  - B_2 [ \dn_{0\pe } (- s(s-1) )]
  - 4\sum_{k=2}^{s-1} (s-k)   
\no \\
&=&    -  \te   {2 \ov 3}    -   4  s (s-1)   \ . \la{007}
 \ee
 In the  conformal 2d  vector $s=1$ case   (corresponding to the Schwinger $\int F \del^{-2} F = \int   A_{m\perp}^2$ action)  we get from  \rf{012}
 $Z_1= \big[\det \dn_{0} ( 0)\big]^{1/2}$ and thus  $B_2^{(1)} = -    {1 \ov 3}$. 
This matches  the expression for $a_s$ \rf{494}   found  from AdS$_3$, 
   in line with   the  $d=4$ and $d=6$  tests of \rf{02},\rf{uh}  discussed above. 

The  $d=2$ CHS  model  discussed  here  is,  of course,  closely related to  spin $s$  W-gravity model   \ci{hul}: 
both have  the same linearized symmetries -- generalized reparametrizations and Weyl  transformations 
for  spin $s$   field. 
The resulting conformal  anomaly coefficient \rf{007}  is indeed equivalent 
 to the    quantum W-gravity  anomaly  given solely  by   the 
   corresponding  bc ghost  contribution to  the 
 central charge $c_{gh} = - 2( 1 + 6s^2 - 6s)$ \ci{hul,pop}.\foot{In standard normalization  with $c=1$  for a real scalar one has 
 $B_2 = { c \ov 24 \pi}  \int d^2x \sqrt g R $ or   $B_2 = { 1 \ov 3} c $ on   $S^2$.}
What   is  remarkable   about  the above derivation  of this result from  the CHS  partition function \rf{012} 
is that it illustrates that  the $d=2$   case, while  somewhat degenerate (having trivial classical action),  
can still be viewed   as a limit of    $d $-dimensional   conformal higher spin theory
(which itself may  then    be  interpreted  as a  natural  $d > 2$   generalization of
 W-gravity).\foot{It may 
be 
 useful  also to  comment on  a special  nature of  the $d=2$ case   regarding 
the structure of induced actions. Starting  with a  matter  Lagrangian coupled to a CHS field 
and  integrating out the matter field one, in general,  gets  a local  logarithmically divergent term  proportional to 
Weyl-invariant  CHS action. 
 In $d =2$ this term   is trivial  which is related to the fact  that  in $d=2$   the  trace anomaly does not contain 
a Weyl-invariant B-type part  \ci{des}  and is  consistent  with  the  vanishing of a  gauge-invariant  CHS action. 
The induced action will contain, of course, also   finite  non-local   terms   which, being anomalous, are  not Weyl-invariant
 and thus are not candidates for  a ``critical'' (i.e. fully symmetric)  CHS action. 
Indeed, the  induced actions  for 
  W-gravity spin $s$ field $\vp_s$  discussed in  \ci{hul,ssn}    may  be written 
  (generalizing the  $s=2$  Polyakov  induced $d=2$  gravity action) as \ 
    $ \int d^2x\   (R_{2s}\,  \del^{-2}   R_{2s}   + ...) $. Here  
  $R_{2s} \sim \del^s \vp_s + ...$   is the  higher spin  curvature, 
  which  is invariant under the generalized  reparametrizations   but not under the generalized Weyl transformations.}   
 

\np 

  \end{document}